\documentclass[draft]{agujournal2019}
\usepackage{url} 
\usepackage{lineno}
\usepackage[inline]{trackchanges} 
\usepackage{soul}
\draftfalse

\journalname{Geophysical Research Letters}

\begin{document}

%
%


\title{Field-Aligned Current Structures during the Terrestrial Magnetosphere's Transformation into Alfvén Wings and Recovery}


%
%




\authors{J. M. H. Beedle\affil{1,2,3}, L.-J. Chen\affil{2}, J. R. Shuster\affil{1}, H. Gurram\affil{2,4}, D. J. Gershman\affil{2}, Y. Chen\affil{5}, R. C. Rice\affil{2,4}, B. L. Burkholder\affil{2,6}, A. S. Ardakani\affil{3}, K. J. Genestreti\affil{7}, R. B. Torbert\affil{1}}

\affiliation{1}{Space Science Center, University of New Hampshire, Durham, NH, USA}
\affiliation{2}{NASA Goddard Space Flight Center, Greenbelt, MD, USA}
\affiliation{3}{Department of Physics, The Catholic University of America, Washington D.C., USA}
\affiliation{4}{Department of Physics, University of Maryland, College Park, MD USA}
\affiliation{5}{Center for Space Physics and Department of Astronomy, Boston University, Boston, MA, USA}
\affiliation{6}{University of Maryland, Baltimore County, MD, USA}
\affiliation{7}{Earth Oceans and Space, Southwest Research Institute, Durham, NH, USA}




\correspondingauthor{Jason M. H. Beedle}{jason.beedle@unh.edu}




\begin{keypoints}
\item On April 24th, 2023, sub-Alfvénic solar wind conditions from a CME caused the Earth's magnetosphere to form an Alfvén Wing formation.
\item The MMS spacecraft observed the dawn-flank wing of the formation, enabling the first in situ measurements of Alfvén Wing current structures.
\item {These current structures were found to be primarily anti-field-aligned, electron-driven, and filamentary.}
\end{keypoints}

%
%

%
%


\begin{abstract}

On April 24th, 2023, a CME event caused the solar wind to become sub-Alfvénic, leading to the development of an Alfvén Wing configuration in the Earth's Magnetosphere. Alfvén Wings have previously been observed as cavities of low flow in Jupiter's magnetosphere, but the observing satellites did not have the ability to directly measure the Alfvén Wings’ current structures. Through in situ measurements made by the Magnetospheric Multiscale (MMS) spacecraft, the April 24th event provides us with the first direct measurements of current structures during an Alfvén Wing configuration. We have found two distinct types of current structures associated with the Alfvén Wing transformation as well as the magnetosphere recovery. {These structures are observed to be significantly more anti-field-aligned and electron-driven than typical magnetopause currents, indicating the disruptions caused to the magnetosphere current system by the Alfvén Wing formation.}

\end{abstract}

\section*{Plain Language Summary}

The solar wind applies pressure on the Earth's magnetic field, distorting it from a dipole into its compressed dayside and stretched tail configuration. However, this typical structure can be disrupted by eruptive solar events such as Coronal Mass Ejections or CMEs, which may cause the solar wind's pressure to drop low enough that it is no longer able to push the magnetosphere back to form a single unified tail. When this occurs, the tail splits into two separate structures, called Alfvén Wings. While this configuration is rare at Earth, it is common at the outer planets and their moons, where Alfvén Wing configurations have been studied and modeled. However, because the observing spacecraft lacked the necessary instrumentation, we have not yet directly observed the Alfvén Wing current structures. On April 24th, 2023, a CME event led to the creation of Alfvén Wing formations in the Earth's magnetosphere. We observed this event using the Magnetospheric Multiscale (MMS) spacecraft, which enabled us to make the first direct observations of Alfvén Wing current structures. These currents were found to be mainly parallel to the local magnetic field, in contrast to typical magnetopause currents, indicating the complex nature of the disrupted magnetosphere's current system.

%
%

%


%
%
%
%

\section{Introduction}

Coronal Mass Ejections (CMEs) are events that project fast moving solar wind plasma and magnetic field lines into the interplanetary medium, creating disturbances and interacting with the background solar wind to create shocks - see \citeA{BeedleRura2022} and sources therein. Upon reaching Earth, these events cause significant disruptions to the magnetosphere's systems through magnetic reconnection on the dayside magnetopause, and the loading of energy into the tail. CMEs are also able to introduce the magnetosphere system to an environment with very low density and high magnetic field strength - e.g. \citeA{Ridley2007}. Because of these low density conditions, CMEs can cause the solar wind to become sub-Alfvénic where the solar wind mach number drops below 1, a significant decrease from the typical solar wind mach number of $\sim$11 \cite{Schunk_Nagy2000}. During such times, as the solar wind pressure collapses, the bow shock may disappear and Alfvén Wings form in the magnetosphere, dividing the once unified magnetotail into a dual configuration such as modeled in the simulations of \citeA{Chane2015} etc. {Note, the Alfvén Wings form above the poles of a body when the IMF is dominantly $B_Z$ orientated, while the wings instead form along the flanks when IMF $B_Y$ becomes more dominant}. 

The current structure associated with the formation of an Alfvén Wing configuration has been modeled and explored in studies for Jupiter, Saturn, and their associated moons - see { \citeA{Neubauer1980}, \citeA{Kivelson2004}}, \citeA{Jia2009} and the associated references - as well as in the Earth's magnetosphere - see \citeA{Ridley2007,Chane2015,Chen2024,Brandon2024}. In both cases, the current structures are theorized to be dominantly field-aligned and run along both edges of the Alfvén Wings, which \citeA{Jia2009} aptly entitle as ``Alfvén Wing currents". {These current structures develop to connect the Alfvén Wing system to the surrounding solar wind flows which enables momentum to be transferred that re-accelerates the slower Alfvén Wing plasma back to the solar wind's speed \cite{Zhang2016}.} The resulting field-aligned current system is then thought to close through perpendicular currents, such as the Pederson currents, in the ionosphere of the moon, or planet in question \cite{Jia2009}. Alfvén Wing formations have also been observed around the Moon by \citeA{Zhang2016} utilizing the ARTEMIS spacecraft. Like the studies in the outer planets, \citeA{Zhang2016} associated the Alfvén Wing structures with field-aligned currents through hybrid simulations. However, no direct in situ measurements of the current structures inside the Alfvén Wing formations were recorded because of the limited instrumental capabilities of the observing spacecraft. 

On April 23rd, 2023, a CME event began impacting the Earth's magnetosphere. This continued into April 24th when the CME's sheath and magnetic cloud were observed passing over the magnetosphere by the Magnetospheric Multiscale (MMS) spacecraft, which were positioned on the dawn-flank of the magnetosphere on an inbound trajectory \cite{LiJenChen2024}. During this time, MMS observed the sub-Alfvénic effects of the solar wind which caused an Alfvén Wing structure to form. Because of MMS's trajectory through this unusual configuration of the Earth's magnetosphere, we were able to study the first in situ measurements of current structures associated with an Alfvén Wing formation. 

\begin{figure}
    \centering
    \includegraphics[width=1\linewidth]{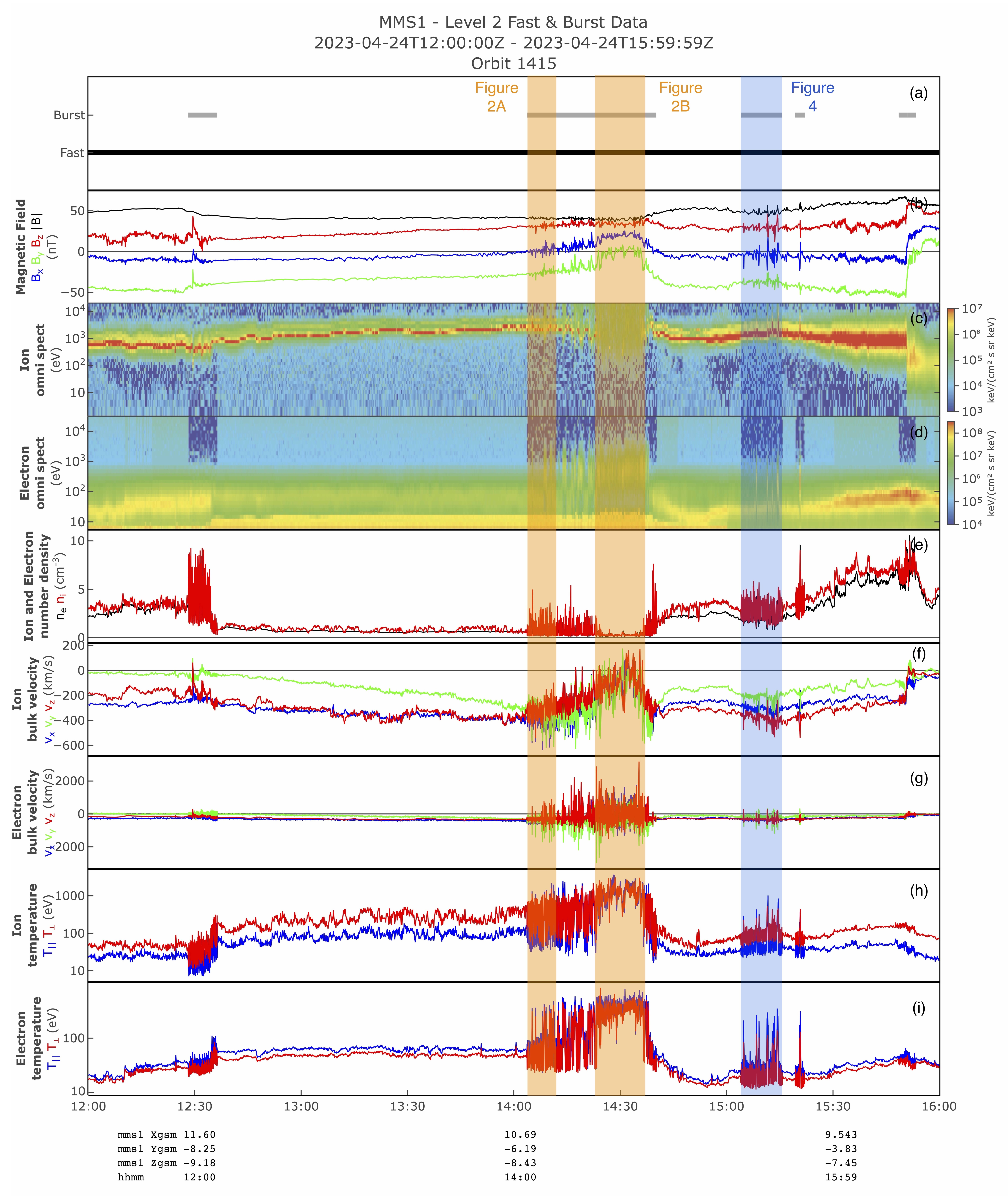}
    \caption{Overview figure from MMS1 of combined fast survey mode data and high resolution burst mode data from 12:00 to 16:00 UT on April 24th. Detailed figures covering select burst mode intervals are included in Figures \ref{fig:Alfven_Wings} and \ref{fig:Recovery}, whose time spans are indicated above by the colored intervals. (a) availability of Burst and Fast mode data across the timeframe, (b) the magnetic field magnitude and components in GSE coordinates, (c)-(d) ion and electron omni directional spectrograms, (e) ion and electron number densities, (f)-(g) ion and electron velocities in GSE coordinates, (h) ion perpendicular and parallel temperature, (i) electron perpendicular and parallel temperature. Note, the position of MMS in GSM coordinates is included at the bottom of the figure. Base figure produced using NASA Goddard's MMS FPI Visualizer software.}
    \label{fig:overview}
\end{figure}

{Figure \ref{fig:overview} provides an overview of MMS 1's fast survey and burst mode data captured during observations made in the unshocked magnetosheath and into the magnetosphere. From panel (b) we can see how the CME was primarily dominated by the IMF's $B_Y$ component, which caused the Alfvén Wings to eventually form on the flanks of the magnetosphere. Panel (e) illustrates the lack of a typical bow shock in the magnetosphere, resulting in a rarefied, unshocked, solar wind-like plasma in the magnetosheath. From approximately 14:00 to 14:40 UT, MMS interacted with the dawn-flank Alfvén Wing and a newly-formed magnetosphere region of closed field lines as is illustrated by the ion deceleration in panel (f) and the ion and electron spectrograms in panels (c) and (d).} After 15:00 UT, the magnetosphere began to recover from the sub-Alfvénic solar wind conditions and gradually reformed its traditional tail configuration \cite{Brandon2024,Chen2024}.

In the proceeding sections, we will cover the current structures observed by the MMS spacecraft from times when MMS's high resolution burst mode data was available: during the Alfvén Wing timeframe from 14:05 - 14:40 UT, and from the Recovery timeframe from 15:05 - 15:15 UT. 

\section{Methods and Data}

For this study, we took data from MMS’s Fast Plasma Investigation (FPI) \cite{FPI} and Fluxgate Magnetometer \cite{FGM} instruments, which enabled observations of the plasma properties and magnetic field conditions from the MMS1 spacecraft. Ion plasma and magnetic field data taken from the MMS spacecraft, and the calculated currents, were interpolated to the 30 ms FPI electron time resolution from the 150 ms ion, and the 10 ms magnetometer time resolutions respectively.

To study the current structures during these time frames, we utilized MMS1's FPI instruments to measure the total current density that arises from the plasma moments, also known as the FPI current:

\begin{equation}
    \textbf{J}_{FPI} = n_e q (\textbf{v}_i - \textbf{v}_e),
\end{equation}

\noindent where $n_e$ is the electron density and $\textbf{v}_i$ and $\textbf{v}_e$ are the ion and electron velocities respectively. We also considered the components of the FPI current parallel and perpendicular to the magnetic field $\textbf{B}$:

\begin{equation}
	{\textbf{J}_{FPI \parallel} = \left( \frac{\textbf{B} \cdot \textbf{J}_{FPI}}{|\textbf{B}|} \right) \hat{\textbf{B}} \ \ , \ \  
	\textbf{J}_{FPI \perp} = \textbf{J}_{FPI} - \textbf{J}_{FPI \parallel}}.
\end{equation}

\noindent The results from the above currents can be seen in Figures \ref{fig:Alfven_Wings} - \ref{fig:Recovery} in subpanels (h)-(j). All quantities are presented in GSE coordinates.

One note regarding the plasma data taken over both the Alfvén Wing and Recovery timeframe, is that the plasma is generally much more rarefied and solar wind-like than is typically expected in the magnetosheath and magnetosphere. Because of the lack of a typical bow shock during this time, the solar wind plasma is no longer shocked, and thus appears as low density and with a high velocity. In order to mitigate the noise that is introduced to the plasma data because of this more solar wind-like plasma, we applied a box-car smoothing algorithm to the plots shown in Figures \ref{fig:Alfven_Wings} - \ref{fig:current_structures}, specifically in subplots (d)-(k) of those figures. 

\section{Alfvén Wing Transformation: 14:00 - 14:40 UT}

\begin{figure}[htbp]
    \centering
    \includegraphics[width=1\textwidth]{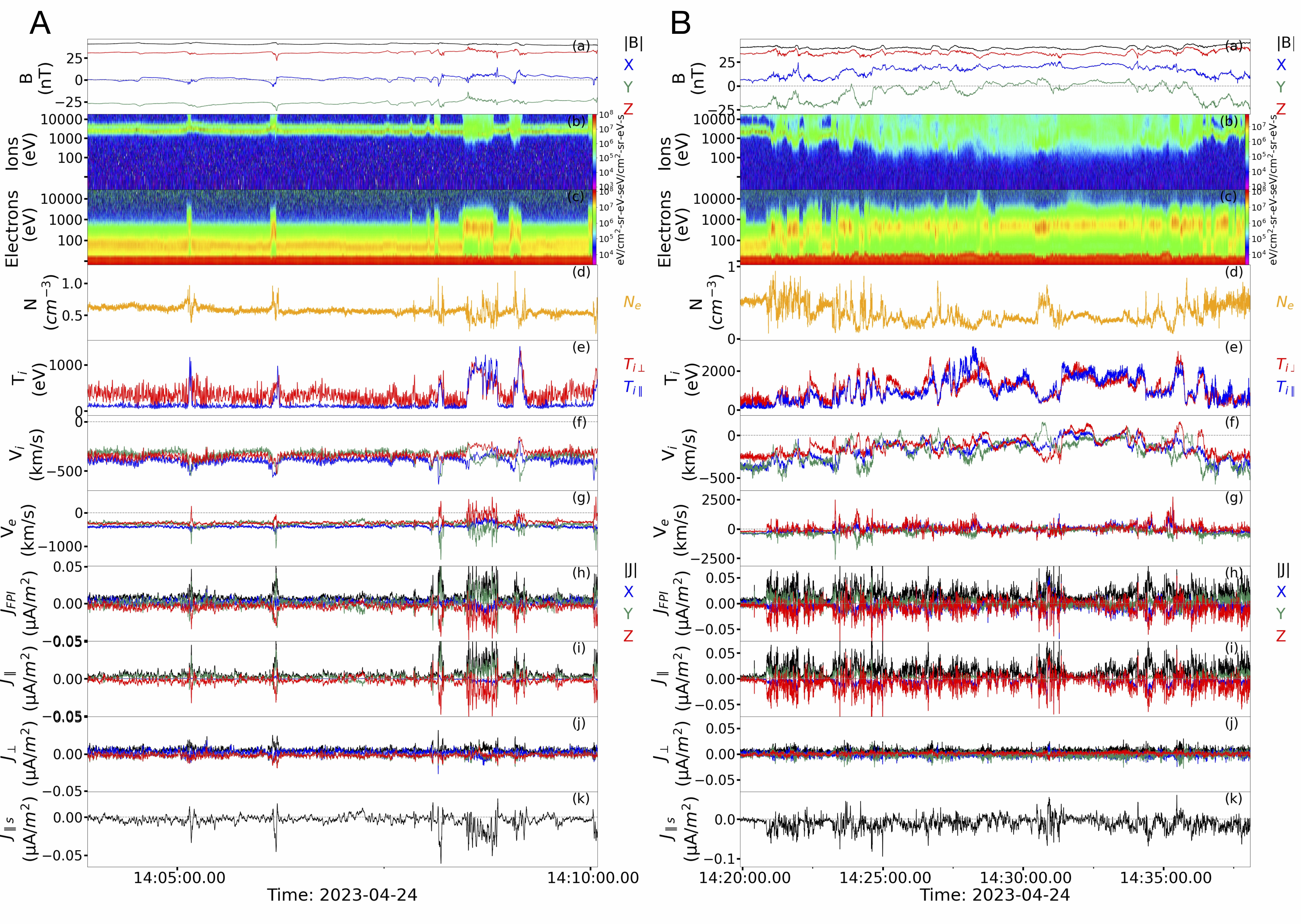}
    \caption{MMS1 burst mode data from 14:05 - 14:10 UT in Panel A when MMS encountered the dawn-flank Alfvén Wing, and 14:25 - 14:35 UT in Panel B when MMS was inside a region of closed magnetic field lines in the Earth's magnetosphere. Subpanels: (a) the magnetic field magnitude and components, (b)-(c) ion and electron omni directional spectrograms, (d) electron number density, (e) ion perpendicular and parallel temperature, (f)-(g) ion and electron velocity, (h) FPI current density, (i) FPI current density parallel to the local magnetic field, (j) perpendicular FPI current density, {(k) scalar parallel current density.} All vector quantities are shown in GSE coordinates.}
    \label{fig:Alfven_Wings}
\end{figure}

{From 14:05 to 14:40 UT on April 24th, the MMS spacecraft encountered the dawn-flank Alfvén Wing and a newly-formed magnetosphere region and made observations of their current structures.} Two intervals from this timeframe are shown in Figure \ref{fig:Alfven_Wings}. 

\subsection{Dawn-Flank Alfvén Wing}

Starting with the 14:05 - 14:10 UT timeframe depicted in Panel A of Figure \ref{fig:Alfven_Wings}, we can see several small-scale current structures associated with the dawn-flank Alfvén Wing. These current structures occur in a hot plasma that is highlighted by increases in the electron energy spectrogram, fluctuations in the electron density, and electron (and to a lesser extent ion) acceleration. {All elements indicating that MMS is encountering magnetospheric origin plasma, which is possibly heated and accelerated by reconnection elsewhere in the magnetosphere.} Taking a look at the current components in subpanels (h)-(j), the current structures are almost entirely field-aligned, especially in the $Y_{GSE}$ and $Z_{GSE}$ directions, {with the scalar parallel current shown in subpanel (k) indicating that the current structures are majority anti-field-aligned.} The perpendicular current density components, on the other hand, barely show current fluctuations above the background. 

An individual current structure at 14:05:10 UT is shown in Figure \ref{fig:current_structures}, Panel A. This zoomed-in view allows us to clearly see how field-aligned the individual structures are, with the perpendicular current density in this case being almost indistinguishable from the background current density fluctuations. This figure also shows how the current structure is primarily reliant on the electron velocity in subpanel (g), indicating these structures also tend to be electron-driven. 

\subsection{Newly-Formed Magnetosphere Region}

Moving to the 14:25 - 14:35 UT timeframe shown in Panel B of Figure  \ref{fig:Alfven_Wings}, MMS is now out of the dawn-flank Alfvén Wing and encountering a region of closed flied lines, such as would occur in a newly produced magnetosphere region, albeit one that sees significant disruption because of the Alfvén Wing transformation. This movement into the magnetosphere is highlighted by the vanishing IMF $B_Y$ component in subpanel (a). 

We can again see current structures during this timeframe, but these are much closer together and also appear inside significantly stronger background fluctuations. Even though the background magnetosphere is disrupted, the current structures during this time are still associated with magnetic field fluctuations and small reversals, but in the $B_y$ component instead of the $B_x$ component as was earlier observed. The ion velocity shows an overall deceleration throughout the timeframe with several reversals throughout 14:25 - 14:35 UT, highlighting the transition of MMS into this magnetospheric region. Just like in the previous panel, the electron velocity tends to dominate over the ion velocity, with the electron velocity showing peaks above 2,000 km/s. Looking closer at the current structures in subpanels (h)-(k), they again appear to be primarily anti-field-aligned and electron-driven like in the dawn-flank Alfvén Wing, but with less defined structure. 




\subsection{Comparison to Typical Magnetosphere Currents}

The current structures recorded during the Alfvén Wing transformation, be it in the dawn-flank Alfvén Wing or magnetosphere region, are found to be primarily field-aligned, electron-driven, and filamentary. This is in contrast to what is typically expected from the magnetopause boundary currents, which are primarily perpendicular to the magnetic field and tend to be ion-driven - e.g. \citeA{Haaland2019,Haaland2020,Beedle2022} etc. {However, these current structures do resemble the simulated and theorized field-aligned ``Alfvén Wing currents" mentioned in \citeA{Neubauer1980}, \citeA{Kivelson2004}, \citeA{Ridley2007}, \citeA{Jia2009}. One interesting note though is that our observed currents are not only field-aligned, but primarily anti-field-aligned as can be seen in subpanel (k) of both timeframes. This raises an interesting question as, from global simulations, MMS is on the sunward facing edge of the dawn-flank Alfvén Wing, which, from these past papers, would imply that MMS should be seeing field-aligned currents, not anti-field-aligned currents. However, as the Alfvén Wings are forming along the flanks of the already complex magnetosphere system, it is possible that their current structures become significantly more complex in this near-Earth environment as will be discussed in Section 5.} 

\begin{figure}[htbp]
    \centering
    \includegraphics[width=1\linewidth]{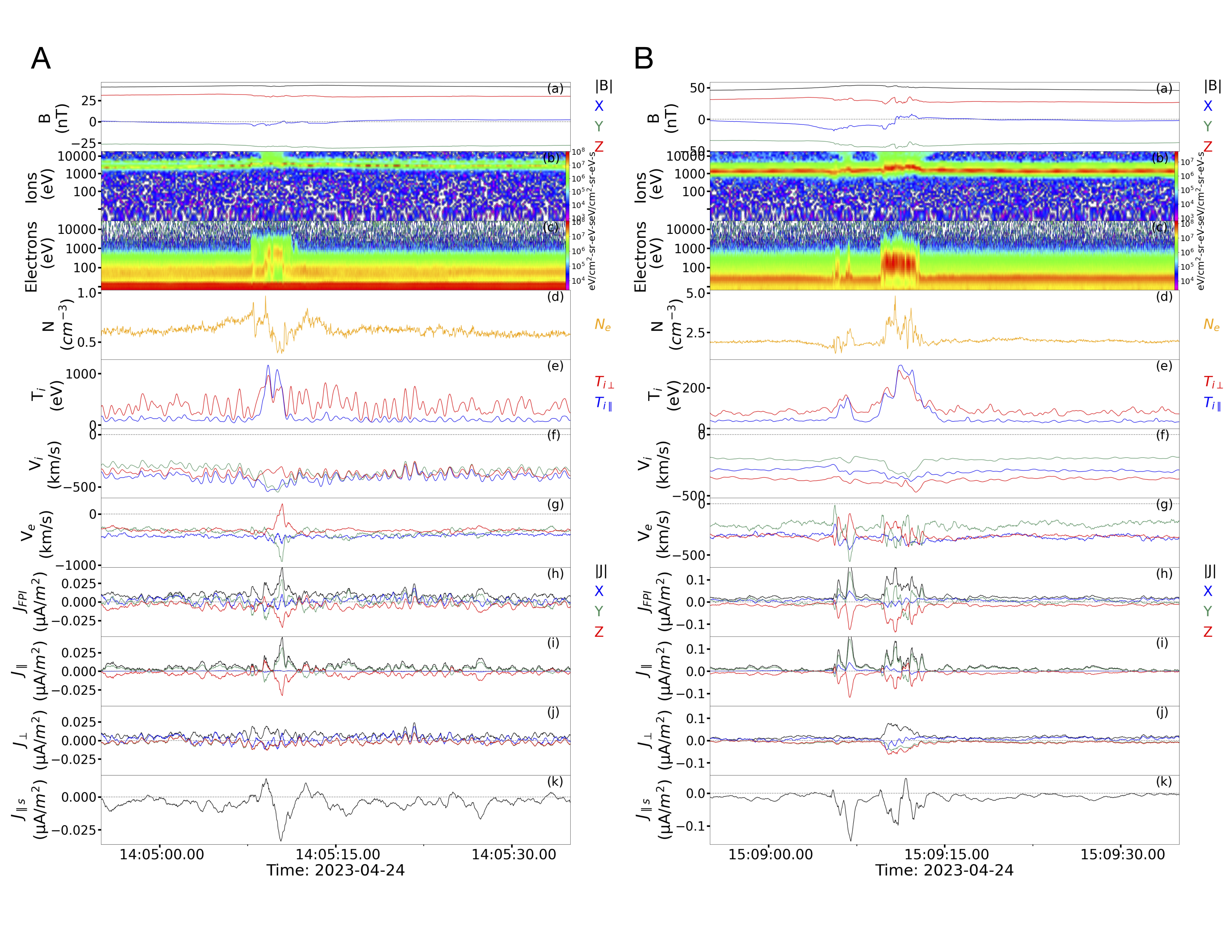}
    \caption{MMS1 burst mode data over two individual current structures from the Alfvén Wing transformation, Panel A, and from the Recovery Phase, Panel B. Subpanels: (a) the magnetic field magnitude and components, (b)-(c) ion and electron omni directional spectrograms, (d) electron number density, (e) ion perpendicular and parallel temperature, (f)-(g) ion and electron velocity, (h) FPI current density, (i) FPI current density parallel to the local magnetic field, (j) perpendicular FPI current density, {(k) scalar parallel current density.} All vector quantities are shown in GSE coordinates. Note the dominance of the anti-field-aligned current density during the Alfvén Wing transformation structure, while the Recovery Phase structure shows comparable parallel and perpendicular current density components.}
    \label{fig:current_structures}
\end{figure}


\newpage

\section{Recovery Phase: 15:05 - 15:15 UT}

\begin{figure}[htbp]
    \centering
    \includegraphics[width=0.9\linewidth]{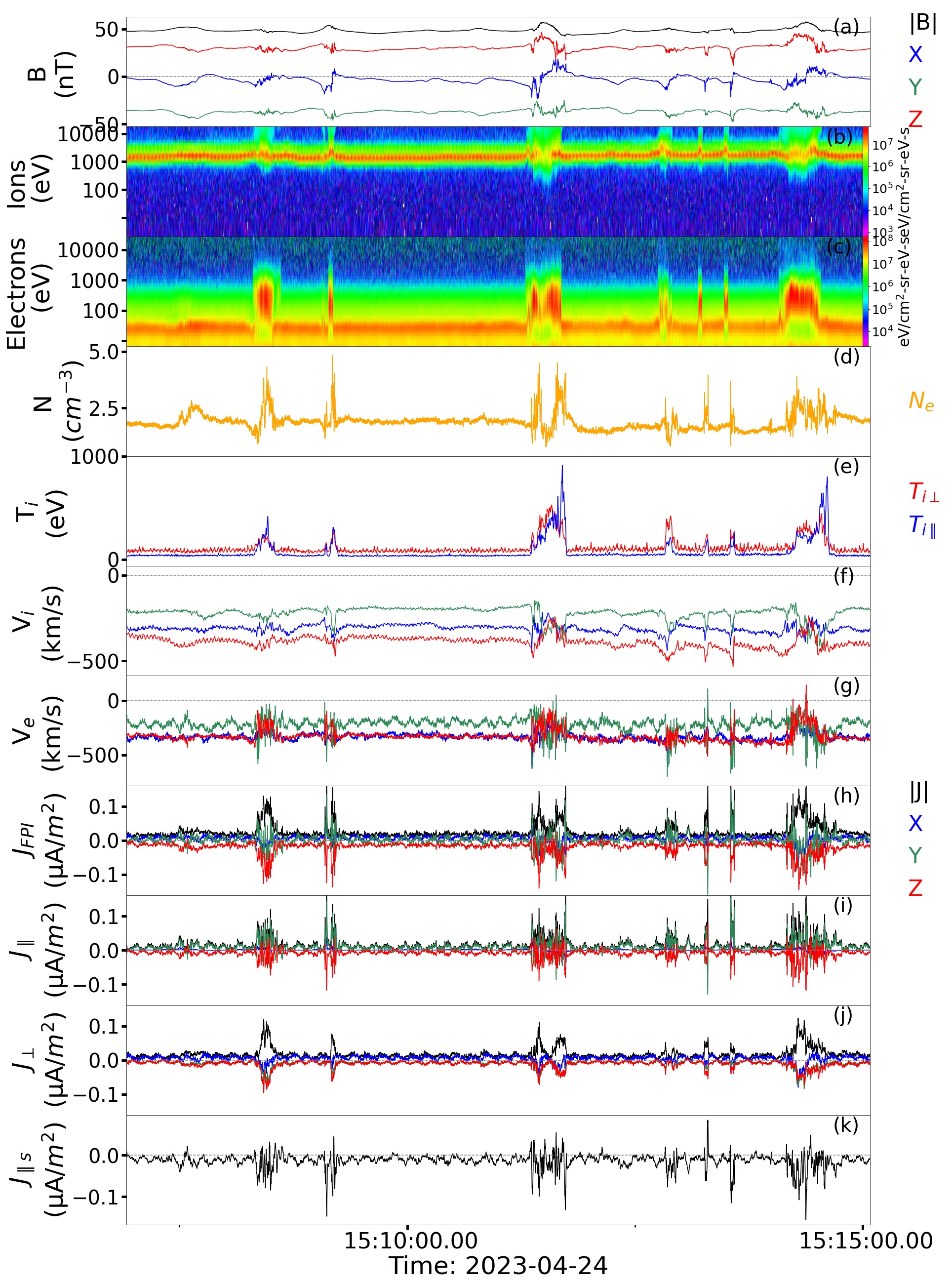}
    \caption{MMS1 burst mode data from 15:05 - 15:15 UT which covers times when MMS was inside the magnetosphere while it was recovering from the Alfvén Wing configuration. Subpanels: (a) the magnetic field magnitude and components, (b)-(c) ion and electron omni directional spectrograms, (d) electron number density, (e) ion perpendicular and parallel temperature, (f)-(g) ion and electron velocity, (h) FPI current density, (i) FPI current density parallel to the local magnetic field, (j) perpendicular FPI current density, {(k) scalar parallel current density.} All vector quantities are shown in GSE coordinates. {Note, the wave-like pattern seen in the $X_{GSE}$ and $Y_{GSE}$ components of the $V_e$ plot is  likely caused by spinetone or spinetone harmonic effects from the FPI instrument - see \citeA{Gershman2019}.}}
    \label{fig:Recovery}
\end{figure}

After 15:00 UT on April 24th, the magnetosphere began its recovery from the Alfvén Wing configuration with MMS recording multiple burst mode intervals during this time. Shown in Figure \ref{fig:Recovery} is one such interval from 15:05 to 15:15 UT where MMS encounters several current structures. These structures occur during fluctuations in the magnetic field with the $B_X$ component showing slight reversals, and the other components showing perturbations. We can also see that the magnitude of the magnetic field tends to increase over the current structures, especially the larger structures near 15:11:15 and 15:15:00 UT. These structures are also associated with sharp increases in the electron spectrograms and electron density, as well as the ion temperature, especially in the parallel component. Both the ion and electron velocities also see significant fluctuations, driving the current density observed in subpanels (h)-(j). Looking at the current density composition, the total current is composed of both significant parallel and perpendicular components, unlike the earlier Alfvén Wing structures. The parallel current density is primarily in the $Y_{GSE}$ and $Z_{GSE}$ directions, while the perpendicular current contains the majority of the $X_{GSE}$ current density with a significant component also along $Z_{GSE}$. Together, these form a total current density with magnitude peaks roughly two times stronger than the Alfvén Wing structures. From these signatures, it appears that MMS is making brief incursions into the now recovering magnetosphere possibly encountering a reforming magnetopause current as is evidenced by the increasingly relevant perpendicular and more ion-driven current.

Figure \ref{fig:current_structures}, Panel B zooms into an individual current structure at 15:09:10 UT where we can again see these characteristics, but in higher detail. Specifically, comparing the ion and electron velocities over the current structure, we can see that both the ion and electron velocities contribute to the formation of the overall current density. This smaller timeframe also illustrates how the Recovery Phase still sees current structures reminiscent of the earlier Alfvén Wing structures as we can see another current structure slightly earlier than the primary one at 15:09:10. This earlier current structure is primarily electron-driven and field-aligned, unlike the 15:09:10 structure which is both ion and electron-driven and has both parallel and perpendicular components. 

Comparing the Recovery Phase structures with the earlier Alfvén Wing structures, we can see several key differences. The first involves the orientation of the current density itself, with the Recovery Phase structures having both significant parallel and perpendicular components, while the Alfvén Wing structures are primarily field-aligned with dominant parallel current components. This indicates that the magnetosphere is, indeed, recovering from the Alfvén Wing configuration with the formerly field-aligned Alfvén Wing current structures beginning to reform into the primarily perpendicular Chapman-Ferraro current structure of the magnetosphere's more typical dayside magnetopause structure - e.g. \citeA{CF1931,Haaland2019,Haaland2020} etc. However, as noted when looking at Figure \ref{fig:current_structures}, Panel B, there are still smaller scale structures in the Recovery Phase that are similar to the earlier Alfvén Wing structures, indicating that the magnetosphere is still in a perturbed state from the Alfvén Wing configuration.

\newpage




\newpage 

\section{{Discussion: The Magnetosphere's Distorted Current System}}

\begin{figure}[htbp]
    \centering
    \includegraphics[width=0.7\linewidth]{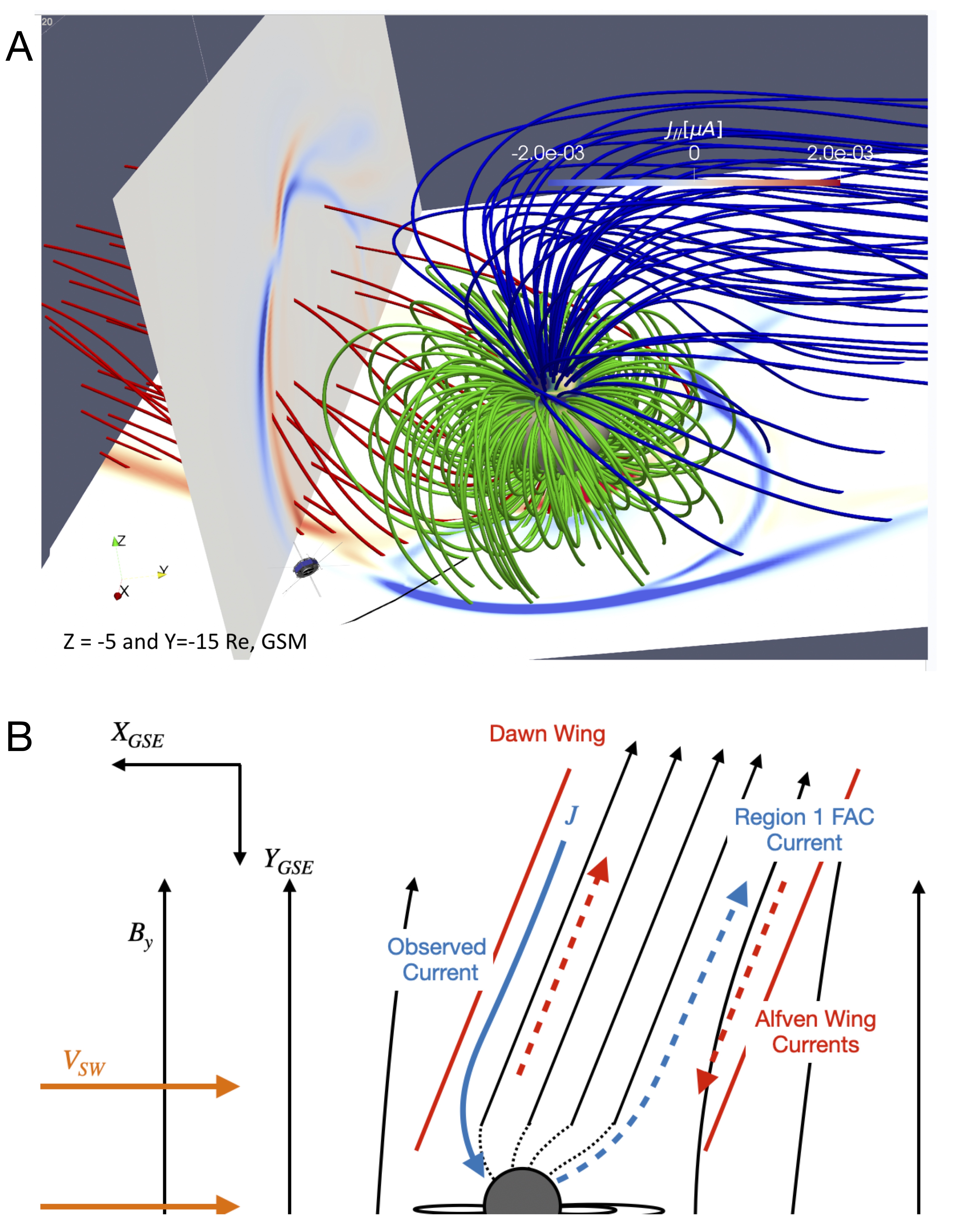}
    \caption{Panel A: Global MHD results from \citeA{Chen2024} of the dawn-flank Alfvén Wing's parallel current density. The dawn-flank field lines are depicted in red, while the dusk-flank field lines are depicted in blue. Closed magnetosphere field lines are shown in green. Panel B: 2D diagram of the dawn-flank Alfvén Wing in Earth's magnetosphere. The sub-Alfvénic solar wind is shown in orange, while the Alfvén Wing structure is indicated in red. The anti-field-aligned current MMS observes on the sunward facing edge of the dawn-flank Alfvén Wing is shown in solid blue, while the unobserved, potentially diverted Region 1 field-aligned current system is shown in dashed blue. The Alfvén Wing currents, as predicted by \citeA{Neubauer1980,Kivelson2004,Jia2009}, are shown in dashed red. Note, the dawn wing field lines connect to the southern hemisphere.}
    \label{fig:Diagram}
\end{figure}

\newpage

{The field-aligned current structures observed by MMS during the Alfvén Wing transformation represent a marked departure from the typical, perpendicular magnetopause current seen at the magnetosphere's edge. However, the magnetosphere itself is no stranger to electron-driven field-aligned currents. The Region 1 and Region 2 field-aligned current systems form connections from the magnetopause, tail, and ring current systems into the Earth's ionosphere - e.g. \citeA{Ganushkina2017}. The Region 1 current system provides current connection to the tail and can become enhanced at the expense the dayside magnetopause's current during periods of geomagnetic activity - see \citeA{Ganushkina2017} and sources therein.}  

{As previously covered, the sub-Alfvénic solar wind from the April 24th, 2024 CME event caused the magnetotail to split into a dual Alfvén Wing formation. In previous theory and simulations, the resulting Alfvén Wing formations show a diverging current system with field aligned currents moving away from the central body along the sunward side of the wings and toward the body on the nightward side \cite{Neubauer1980,Kivelson2004,Jia2009}. This Alfvén Wing current system is shown denoted in blue in Panel B, Figure \ref{fig:Diagram}. However, unlike the majority of these previous cases, the Earth's Alfvén Wings do not form around a simple conducting body, but in an already complex and fully formed magnetosphere with its own current circulation. While the formation of the Alfvén Wings would doubtless cause an unprecedented disruption to the usual flow of the magnetosphere's currents, it is likely that, instead of severing the current system's flow, it is diverted into the Alfvén Wing formations themselves, replacing the Region 1's field-aligned current connection from the tail to the ionosphere with a connection from the wings into the ionosphere. An example of this more complex field-aligned current system can be seen in the global MHD models of \citeA{Chen2024} as shown in Panel A of Figure \ref{fig:Diagram}, which depicts a nested field-aligned current system with both anti-field-aligned and field-aligned components. One explanation of this more complex nested nature involves the reformation of the field-aligned Region 1 current system as is shown in blue in Panel B, Figure \ref{fig:Diagram}, which connects the wing current structure into the ionosphere. This current system would include a sunward-side, electron driven anti-field-aligned current, which matches with MMS's observations and the global MHD model in Panel A. This more complex structure would also explain why MMS continues to see such field-aligned structures even when passing from interactions with the dawn-flank Alfvén Wing into the newly formed magnetosphere region, and eventually into the Recovery Phase magnetosphere, as the diverted Region 1 currents would persist from the wing structure into the magnetosphere. However, given MMS's observations are only on the sunward edge of the dawn-flank Alfvén Wing, a more solid picture of this current system complexity, including the potentially nested current structure, is unable to be seen solely through these in situ observations.}

\section{Overview and Conclusions}

The April 24th CME event provided an unprecedented opportunity to observe Alfvén Wing structures in the Earth's magnetosphere. MMS's position at the base of the dawn-flank Alfvén Wing allows us to present direct measurements of the current structures associated with an Alfvén Wing for the first time. From this analysis, we found that the Alfvén Wing current structures are primarily anti-field-aligned and electron-driven, in contrast to the typical magnetopause/magnetosphere current system. This finding resembles previous expectations of the Alfvén Wing currents as were simulated and predicted in \citeA{Neubauer1980,Jia2009,Chane2015} etc. {However, unlike these previous studies, MMS's observations reveals anti-field-aligned current structures that persists from the dawn-flank Alfvén Wing into a newly-formed magnetosphere region, which suggests a significantly more complex current structure caused by Earth's dynamic magnetosphere.} 

MMS also recorded data during the Recovery Phase of the magnetosphere as the driving conditions relaxed, and the magnetosphere started to recover its more typical structure. During this Recovery Phase, MMS recorded brief incursions into the magnetosphere and observed current structures with significant parallel and perpendicular components, indicating that the magnetopause current system was starting to recover its usual circulation. While none of the current structures presented in this paper directly suggest magnetic reconnection occurred at the site of MMS's observations, the presence of a freshly formed magnetosphere region in the 14:25 - 14:35 UT timeframe indicates that reconnection was active elsewhere in the Alfvén Wing structure. The topic of magnetic reconnection, and its impacts on the magnetosphere during the Alfvén Wing transformation, will be covered in subsequent papers on this event.

\newpage

\section{Open Research}

The MMS data used in this study is publicly available from the FPI and FIELDS datasets provided at the MMS Science Data Center, Laboratory for Atmospheric and Space Physics (LASP), University of Colorado Boulder \cite{MMS_dataset} and from the following MMS1 datasets: \citeA{MMS1FPI_i,MMS1FPI_e,MMS1FIE}.


\acknowledgments
We thank the MMS team and instrument leads for the data access and support. We also thank the FPI Vizualizer team for their support and data visualization tools, made publicly available at https://fpi.gsfc.nasa.gov/. Additionally, we thank the pySPEDAS team for their support and data analysis tools. This research was supported by the NASA Magnetospheric Multiscale Mission in association with NASA contract NNG04EB99C. J. M. H. B. was supported through the cooperative agreement 80NSSC21M0180, and through the MMS grant 14N820-UZSPRT.


%
%




\bibliography{main.bib}

%
%
%
%
%

\end{document}